\documentclass[prl,aps,twocolumn,showpacs,showkeys,superscriptaddress,groupedaddress]{revtex4}  
\usepackage{graphicx}  
\usepackage{dcolumn}   
\usepackage{bm}        
\usepackage{amssymb}   
\usepackage{color}

\begin{document}

\title{Resolving the $\Delta(1232)$ partial width anomaly: Complex pole residue is not a fundamental resonance property}

\author{S.~Ceci}
\email{sasa.ceci@irb.hr}
\affiliation{Rudjer Bo\v{s}kovi\'{c} Institute, Bijeni\v{c}ka  54, HR-10000 Zagreb, Croatia}
\author{H.~Osmanović}
\affiliation{University of Tuzla, Univerzitetska 4, 75000, Tuzla, Bosnia and Herzegovina}
\author{B.~Zauner}
\affiliation{Institute for Medical Research and Occupational Health, Ksaverska 2, HR-10000 Zagreb, Croatia}

\begin{abstract}
The resonant properties of excited hadrons are commonly identified with the complex pole positions and residues of the scattering amplitude. The mass and total decay width are given by position, whereas the partial width is given by the magnitude of the residue. If this identification was correct, the partial width of famous $\Delta(1232)$ would be larger than its total width. By using a simple model that predicts residue phases of prominent baryons $N^*$, $\Delta$, $\Lambda$, $\Sigma$, and low mass mesons, we resolve this anomaly and show that the residue cannot be a fundamental resonant property.
\end{abstract}


\maketitle

In 1936, in their study of neutron scattering, Breit and Wigner introduced their notable approximate formula for the resonant scattering amplitude \cite{BreitWigner1936}. In it, the resonance was parameterized by the real and imaginary part of the amplitude pole, as directly related to the mass and the total decay width, respectively. In addition, the magnitude of the complex pole residue was related to the partial decay width. In 1979, in his study of pion-nucleon scattering Cutkosky introduced an additional resonant parameter, the complex pole residue phase \cite{Cutkosky1979,Cutkosky1980}. Unlike the first three resonant parameters, this one did not have obvious physical meaning. H\"ohler used a unitary addition of resonant and background terms \cite{Hohler1993, HohlerBible}, and the pole residue phase turned out to be equal to twice of the background complex phase. This raises a natural question: How can residue be a fundamental resonant parameter if its phase depends on the background? 

Following Cutkosky's results, a similar model was used with data from $\pi N \rightarrow \eta N$ scattering done by Batinić \cite{Batinic1995, Batinic2010} (Zagreb model) and Vrana \cite{Vrana2000}. Alternatively, advanced K-matrix analyses are performed and all pole parameters are extracted within the George Washington University model \cite{Arndt2006}, the Bonn-Gatchina model \cite{Anisovich2012} followed by the CBELSA/TABS collaboration analysis \cite{Sokhoyan2015}, and by the J\"ulich model \cite{Ronchen2015, Ronchen2022}. All these analyses are providing increasingly better results with overall smaller uncertainties and (hopefully) smaller systematic ones. This revealed another intriguing problem with the residue. Assuming the Breit-Wigner interpretation of a residue magnitude as a partial decay half-width, it turns out that for the best known nucleon excited state, $\Delta(1232)$, the partial width to $\pi N$ channel is larger than the total decay width. The difference is small, but statistically significant. 

We note here that from the beginning of the nucleon resonance research, the pole parameters were not the only resonant parameters in this field. Both Cutkosky \cite{Cutkosky1980} and H\"ohler \cite{Hohler1993,Hohler1979,HohlerBible} used the alternative ones, called simply the conventional resonant parameters. Notable results extracted from these parameters are achieved in the Kent State model \cite{Manley1992, Manley1995, Shrestha2012, Hunt2019}. In the tables of the {\it Review of Particle Properties} \cite{PDG} these parameters are, somewhat unfortunately, dubbed the Breit-Wigner parameters: mass, width, and the branching fraction. In single channel situations, they are sometimes obtained by a Flatt\'e-like formula \cite{Flatte1976} which assumes more realistic energy dependent decay width. In simple single-resonance cases, the Breit-Wigner mass is equal to the zero of the real part of the scattering amplitude. However, there is no model-independent definition for these parameters in cases with more resonances and channels. In all definitions, the partial width is smaller than, or equal to the total decay width, and there is no parameter analogous to the residue phase. 

There are other processes that show similar resonant behavior. The kaon-nucleon scattering is studied by the Osaka-Argonne group \cite{Kamano2015} and, again, the Bonn-Gatchina group \cite{Sarantsev2019}. They produce both pole and Breit-Wigner parameters. In the pion-pion scattering, the residue phase of meson resonances is, unfortunately, seldom reported. However, there are results by Hoferichter \cite{Hoferichter2024} for $f_0(500)$, $\rho$ meson, $f_0(980)$, and $f_2(1270)$. 

In this Letter we show that the complex pole residue of a resonance is not its fundamental property. We use a minimally improved version of the Breit-Wigner formula \cite{Ceci2017}, which includes the reaction threshold in addition to the resonant pole, to confirm that it correctly predicts residue phases for most of the first (and most prominent) resonances seen in $\pi \pi$, $\pi N$, and $K N$ scattering ({\it i.e.}, $N^*$, $\Delta$, $\Lambda$, and $\Sigma$ baryons). Then we estimate the residue phase of $f_0(980)$, a second meson resonance in $S_0$ partial wave of $\pi \pi$ scattering, using a unitary approximation based on H\"ohler's unitary addition, and similar to the one used in \cite{Ceci2017}. Finally, we go back and use the same approximation on the $\pi N$ $P_{33}$ partial wave, to show that it produces the magnitude of the complex pole residue of $\Delta(1232)$ such that the partial decay width is larger than the total one. Exactly as was noted in all analyzes of this famous resonance. The residue is a property of amplitude, not resonance.  


To estimate a dimensionless elastic amplitude $T$, as defined in Ref.~\cite{HohlerBible}, we use an improved version of the Breit-Wigner formula \cite{Ceci2017} 
\begin{equation}
    T=x\,e^{i\,(\delta_R+\beta)}\,\sin   (\delta_R+\alpha).
\end{equation}
Here, the resonant phase shift $\delta_R$ is defined as 
\begin{equation}
    \tan \delta_R = \frac{\Gamma/2}{M-E},
\end{equation}
which produces amplitude $T$ having a simple pole at the position of \mbox{$M-i\,\Gamma/2$} in complex plane of the center of mass energy $E$. If $\alpha$ and $\beta$ are both set to zero, this form turns out to be the original (yet non-relativistic) Breit-Wigner formula for the resonant amplitude. 

Phase $\alpha$ is introduced in Ref.~\cite{Ceci2013} (where it was called $\delta_0$) to describe the asymmetry of cross-section peaks. There, amplitude $T$ at the elastic threshold energy $E_{0}$ is approximated to have a simple zero, and tangent of $\alpha$ was given by 
\begin{equation}
    \tan \alpha = \frac{\Gamma/2}{M-E_{0}}.
    \label{eq:alpha}
\end{equation}

Phase $\beta$ has the value of the phase shift background close to the resonance, and brings about the shift between real part of the pole position $M$ and the Breit-Wigner mass $M_{BW}$ 
\begin{equation}
    \tan \beta = \frac{M-M_{BW}}{\Gamma/2}.
    \label{eq:beta}
\end{equation}

The elastic residue phase $\theta$ is, within this approximation, given by
\begin{equation}
    \theta = \alpha + \beta.
    \label{eq:theta}
\end{equation}

\begin{figure}[h]
    \centering
    \includegraphics[width=0.47\textwidth]{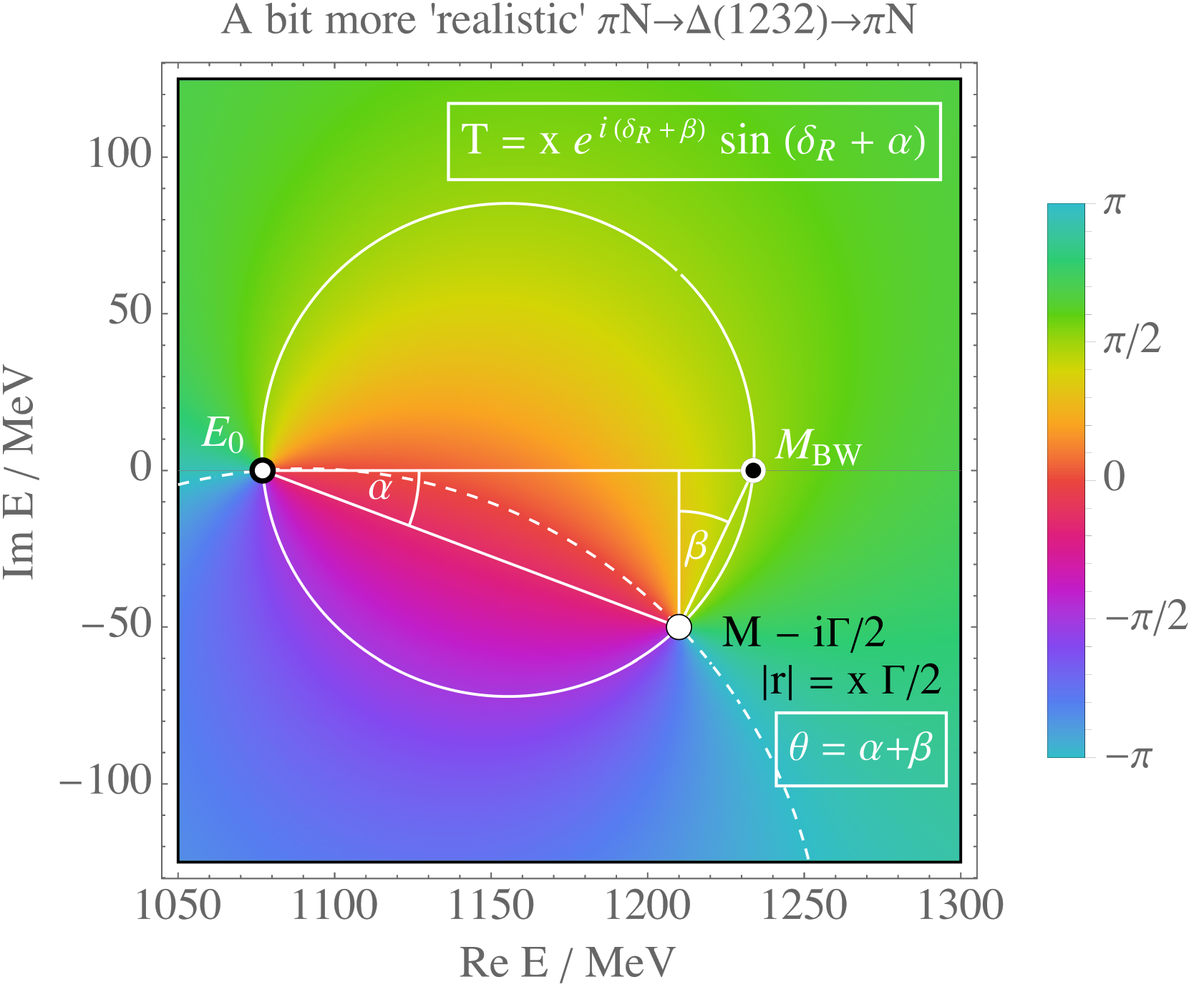}
    \caption{Illustration of our model. Colors are complex phase of amplitude $T$. Full circle represents pure imaginary amplitude, and dashed line is pure real amplitude.  Numerical values are taken from the $\Delta(1232)$ case}
    \label{fig:model}
\end{figure}

The remaining model parameter is the elastic branching fraction $x$. The complex pole elastic residue magnitude equals to $x\, \Gamma/2$, which is the partial decay half-width in the original Breit-Wigner sense. All these model elements are illustrated in Fig.~\ref{fig:model}.

\begin{figure}[h]
    \centering
    \includegraphics[width=0.23\textwidth]{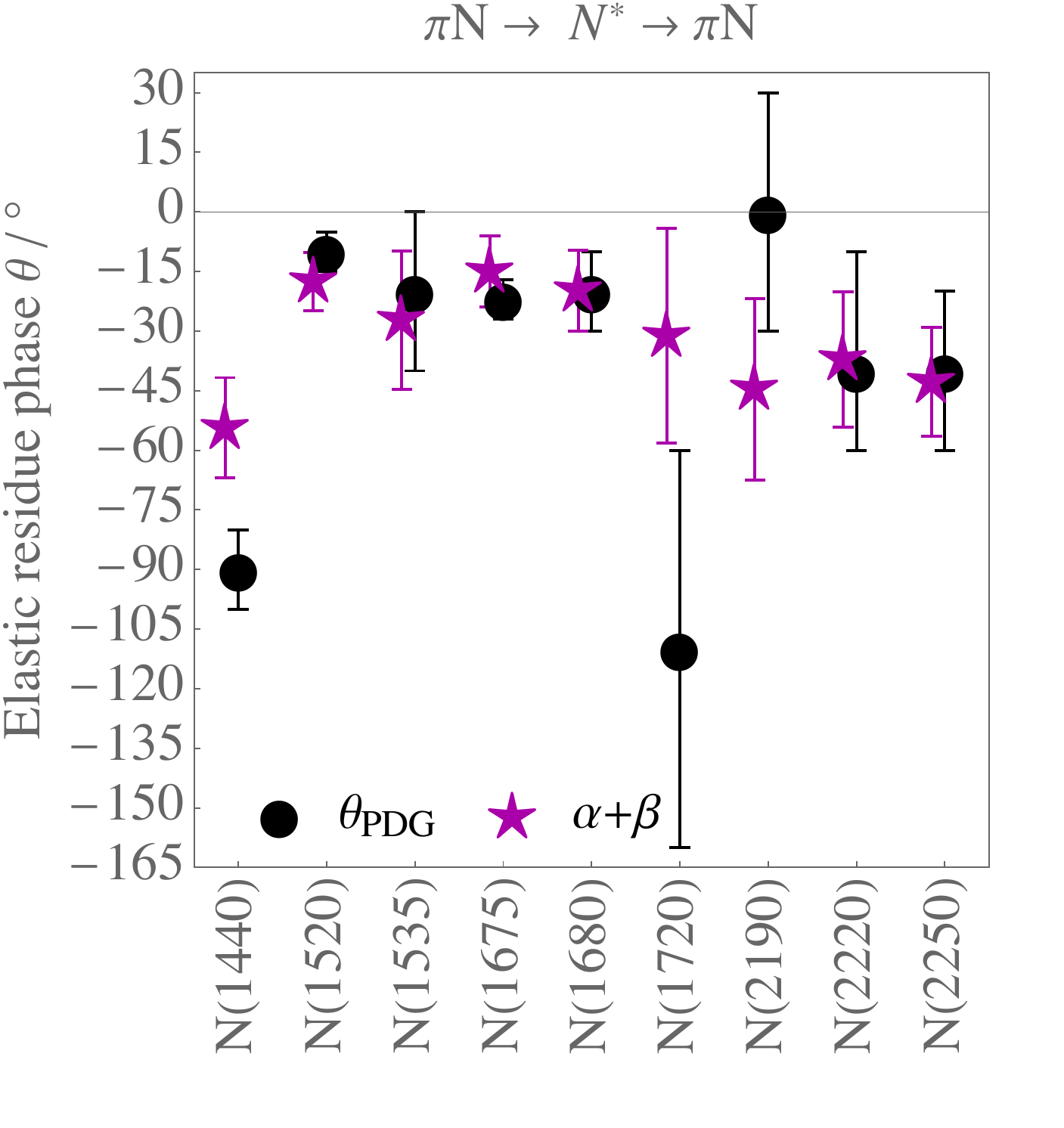}
    \includegraphics[width=0.23\textwidth]{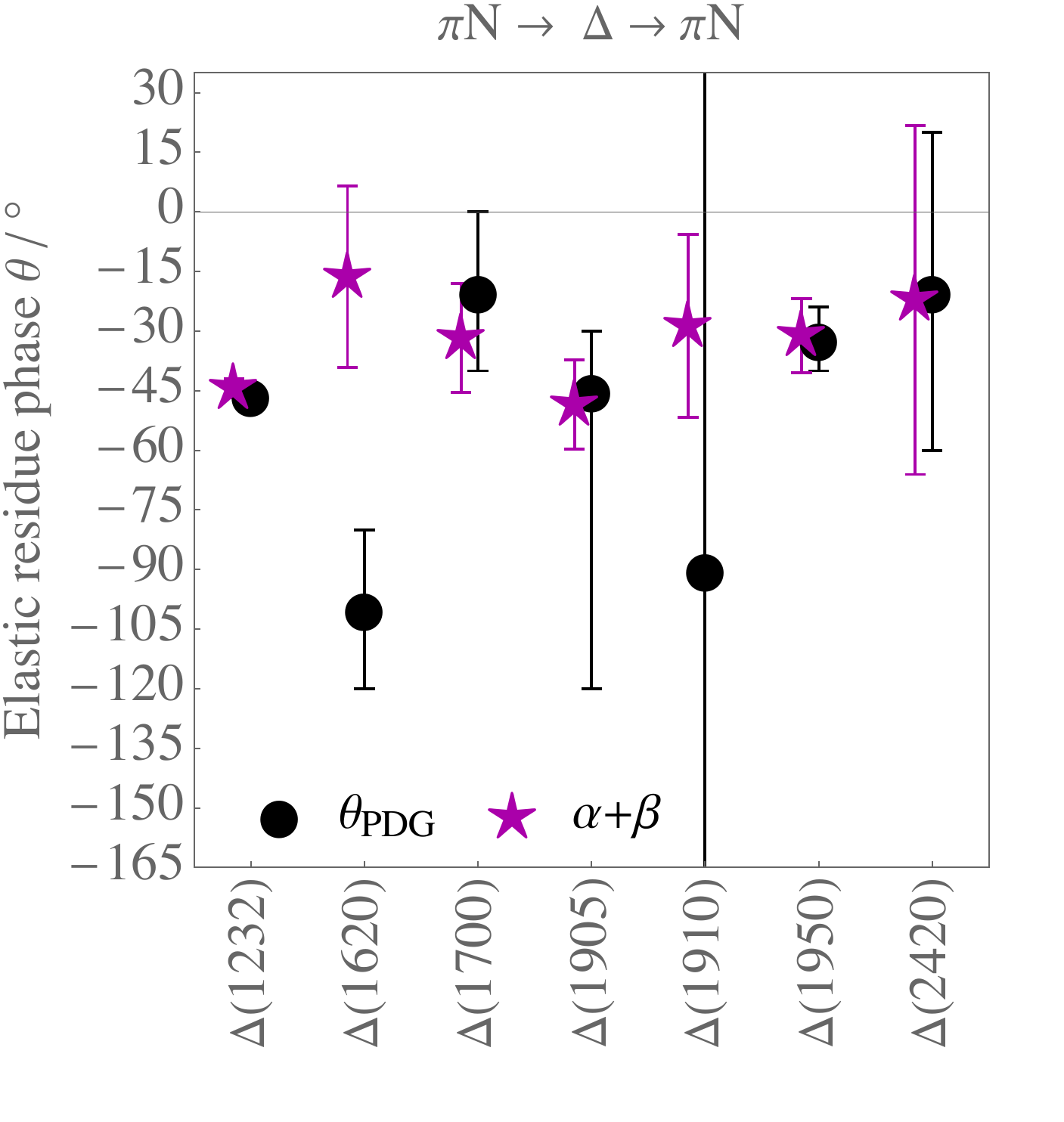}
    \includegraphics[width=0.23\textwidth]{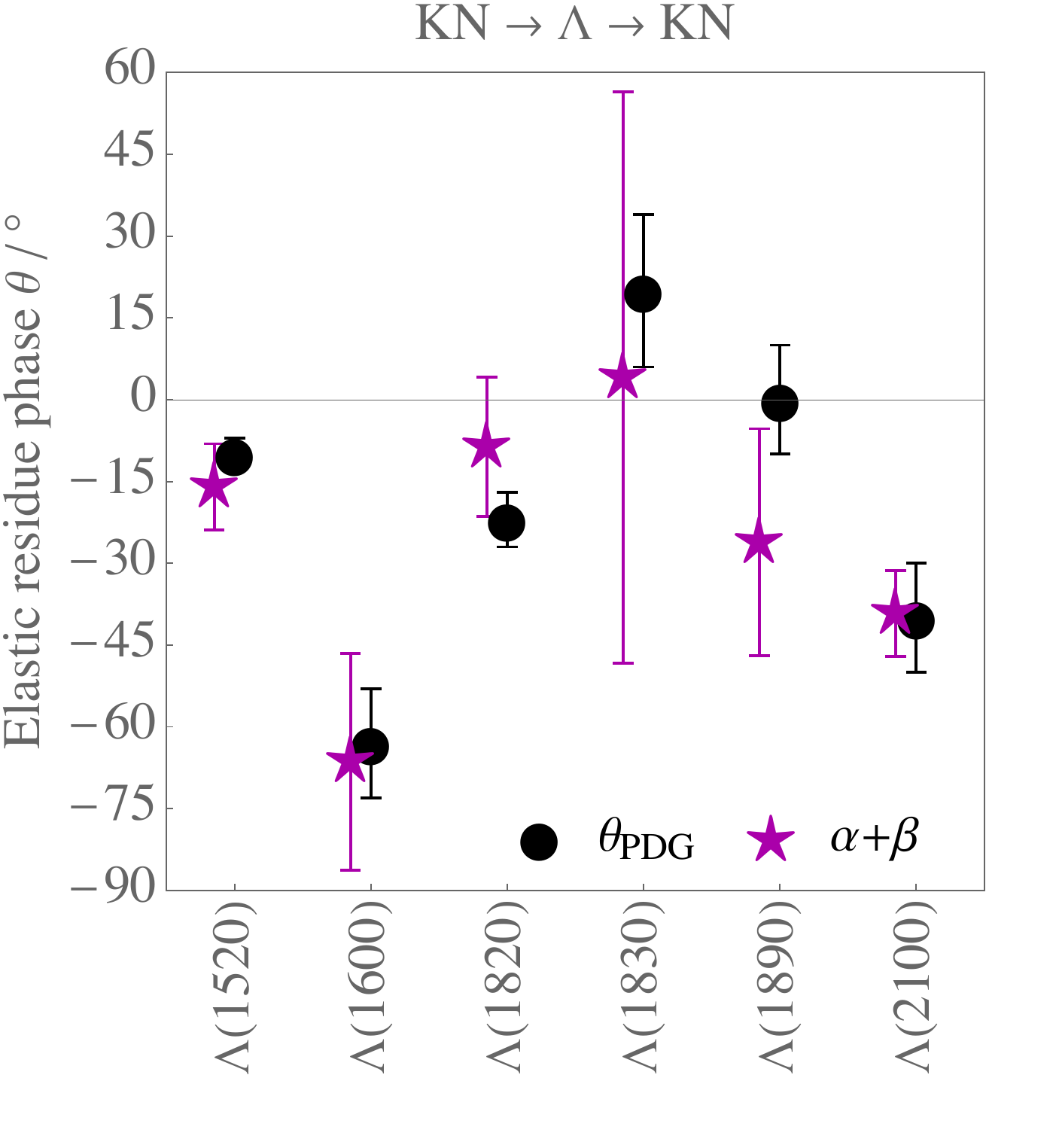}
    \includegraphics[width=0.23\textwidth]{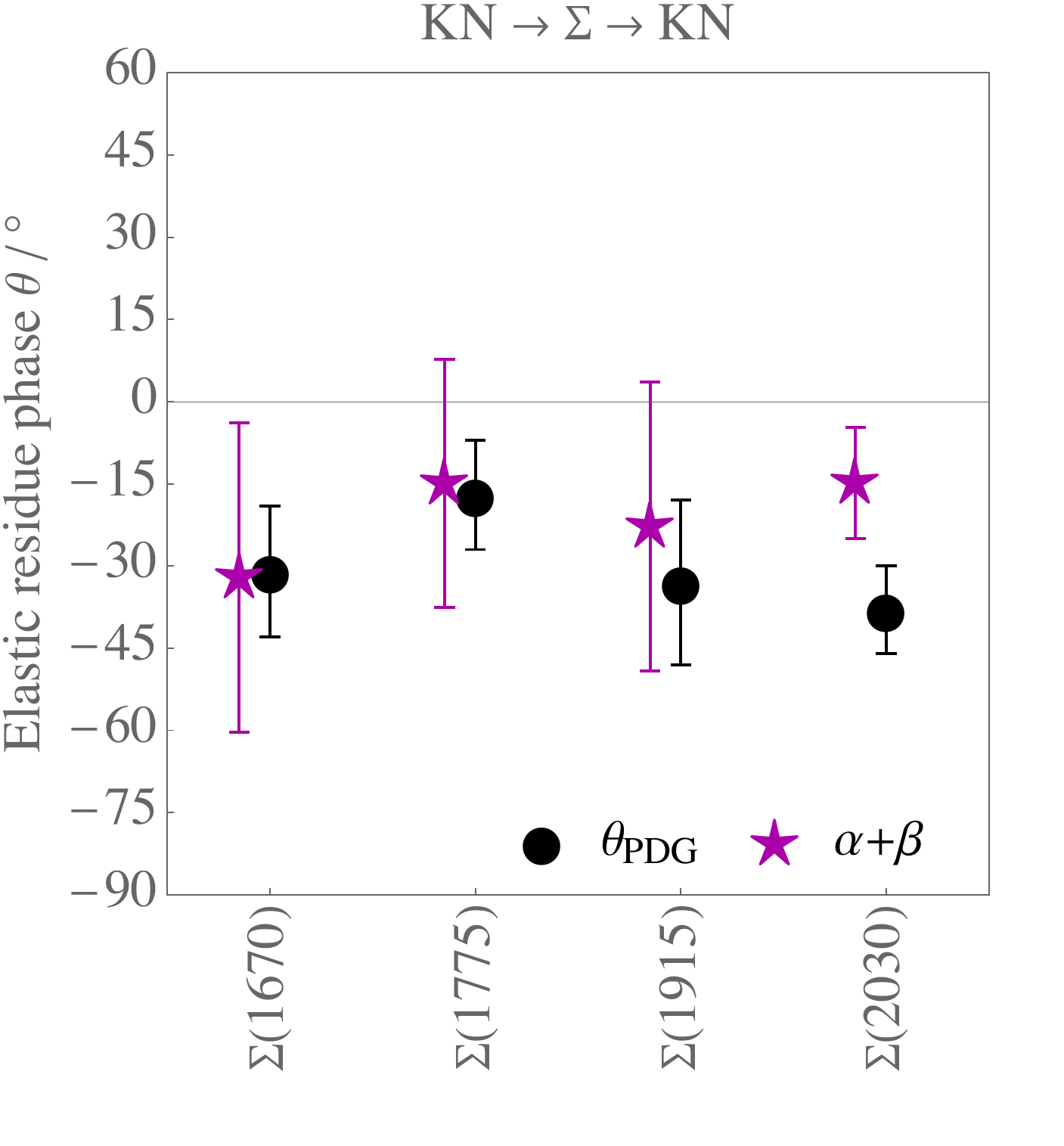}
    \caption{The elastic residue phase for the first resonance in each partial wave in the $\pi N$ and $KN$ elastic scattering. Circles are PDG estimates \cite{PDG}, stars are results of our model. Roughly 2/3 are in perfect agreement with the PDG values }
    \label{fig:baryons}
\end{figure}

We apply this model on the well established lowest mass $N^*$, $\Delta$, $\Lambda$, and $\Sigma$ resonances in each partial wave seen in $\pi N$ and $KN$ elastic scattering. Input parameters are the averages and error estimates of the pole positions and Breit-Wigner masses taken from the year 2024 edition of the {\it Review of particle physics} \cite{PDG}. We calculate $\alpha$ for each resonance from known elastic threshold ($\pi N$ or $KN$), and the pole position using Eq.~(\ref{eq:alpha}). Then we calculate $\beta$ from the known Breit-Wigner mass, and the pole position using Eq.~(\ref{eq:beta}). The elastic residue phase is then the sum of the two, as given in Eq.~(\ref{eq:theta}). Error estimates of the results are calculated using a standard error propagation formula, assuming all errors are independent. If there were different upper and lower errors, we have used the larger one. The final results are shown in Fig.~\ref{fig:baryons}. 

Evidently, the model works rather well for roughly 2/3 of the particles we analyzed. That is quite surprising, because it is extremely simple. The most mind boggling thing here is that this model completely ignores the threshold behavior of the amplitudes. There should be a momentum-like (square root type) branch point with strong angular-momentum related effects of centrifugal barrier. Instead, a simple zero at the threshold somehow worked out. This unexpected feature was first noted on experimental cross sections in Ref.~\cite{Ceci2013}, and then confirmed for excited nucleons in Ref.~\cite{Ceci2017}.    

For the remaining 1/3 of resonances for which the model failed to reproduce the measured $\theta$, it most often fails by a substantial amount. This drastic departure from a typical behavior is sometimes related to a strong coupling of the resonance to a channel that opens in its close proximity. In the case of $N(1440)$, the famous Roper resonance, we know that there is a $\pi \Delta$ channel opening almost at the pole mass, and that the resonance strongly couples to that channel (branching fraction of 20-22\%, see e.g., Refs. \cite{Anisovich2012,Sokhoyan2015,Hunt2019}). This way, on the real axis we do not see one resonance pole only, but rather feel the presence of both the resonant and the shadow pole. This is why such a strong discrepancy is expected.

It is important to stress that we have similar shadow pole situation for each new channel that opens. However, if the resonance is not strongly coupled to this new channel, i.e., the resonance's decay probability to that particular channel is low, the shadow pole will be roughly at the same place, and have other parameters the same as the elastic pole. Basically, everything will be as if there was no new channel. That is the reason why even the most simple resonant models are, at least sometimes, known to work rather well. 

The residue phase in the meson sector is seldom extracted. In $\pi\pi$ elastic scattering there are four resonances that posses all parameters necessary for our analysis: $f_0(500)$, $\rho(770)$, $f_0(980)$ and $f_2(1270)$. Since $f_0(500)$ and $f_0(980)$ have the same quantum numbers, we had to add them in a unitary way. To do it, we use the approximation from Ref.~\cite{Ceci2017}. 

This is what has to be done. First we note that our amplitude $T$ is just one element of the full $T$-matrix. The scattering matrix $S$ is related to the $T$-matrix as $S_{ij}=\delta_{ij}+2\,i\,T_{ij}$. Full $S$-matrix is unitary, and it can be produced by multiplying unitary $S$-matrix contributions of each resonance and background, as in Refs.~\cite{HohlerBible, Manley1995}. Yet, it is extremely difficult, if not impossible, to know all of the $S$-matrix elements. Lacking the better option, in Ref.~\cite{Ceci2017} it was assumed that the dominant part of the diagonal $S$-matrix element, for the elastic process that we analyze, will be the product of the same diagonal elements of all single-resonance $S$-matrices. This boils down to 
\begin{equation}
  1+2\,i\,T\,\approx\,
  \prod_{r} \left(1+2\,i\,T^{(r)}\right). \label{eq:UnitaryAddition}
\end{equation}
If there was only one channel, and $S$-matrix was scalar, this formula would have been equivalent to H\"ohler's unitary addition \cite{HohlerBible}. 

\begin{figure}[h]
    \centering
    \includegraphics[width=0.47\textwidth]{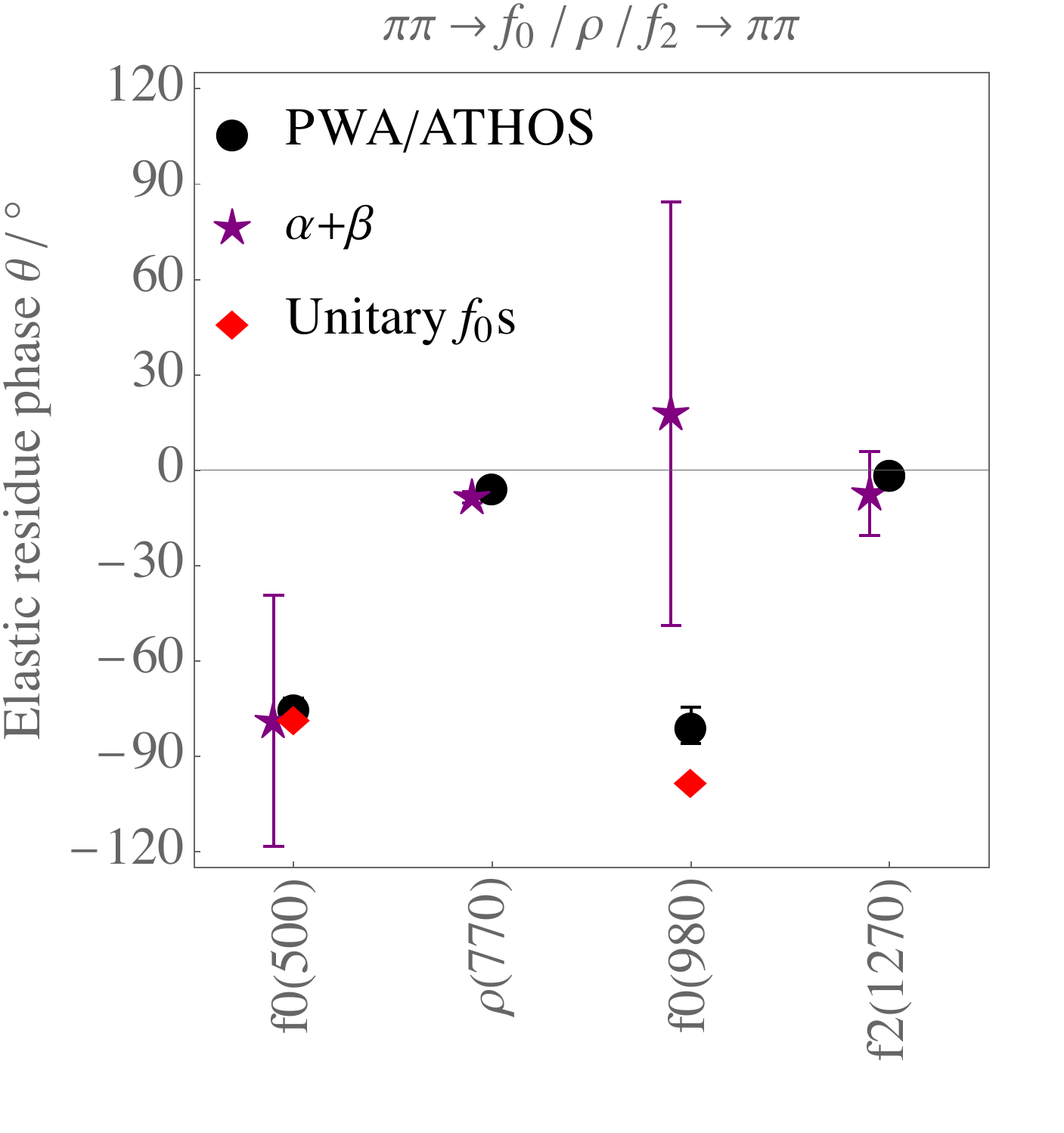}
    \caption{The elastic residue phase for the meson resonances in the $\pi \pi$ elastic scattering (using the same convention as in hadrons). Diamonds are obtained by using Eq.(\ref{eq:UnitaryAddition}) for $f_0(500)$ and $f_0(980)$ }
    \label{fig:mesons}
\end{figure}

\begin{figure}[h]
    \centering
    \includegraphics[width=0.47\textwidth]{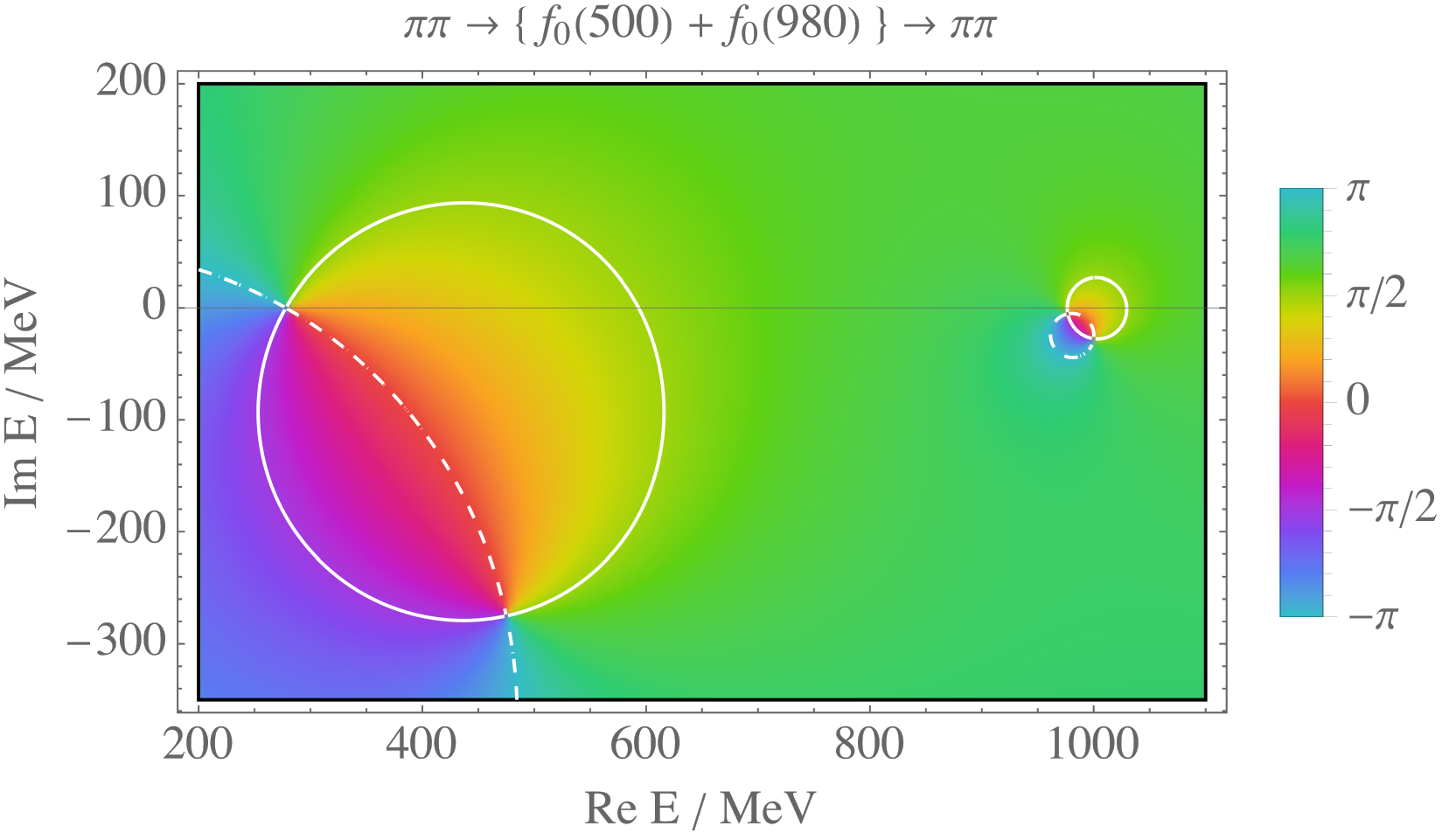}
    \caption{The complex amplitude phase for a unitary combination of $f_0(500)$ and $f(980)$ }
    \label{fig:f0mesons}
\end{figure}

Next, we use it to calculate the residue phase of $f_0(500)$ and $f_0(980)$. For mesons we use the non-relativistic Breit-Wigner parametrization, as is usually done for the excited baryons, which changes the value of meson resonance residue phases. The results are given if Fig.~\ref{fig:mesons} and Fig.~\ref{fig:f0mesons}. Circles are the empirical phases from Ref.~\cite{Hoferichter2024}, stars are results that assume the resonance is the only one in the partial wave, and diamonds are the two meson resonances in $S_0$ partial wave combined using Eq.~(\ref{eq:UnitaryAddition}). 

Since this approximation worked quite well on a $f_0(500)$ and $f_0(980)$ pair, we try it on a famous $\Delta(1232)$, and the other resonances sharing the same quantum numbers, $\Delta(1600)$ and $\Delta(1920)$. Here, we wanted to see how will the presence of additional resonances affect $\Delta(1232)$. 

The problem we are addressing is the fact that all relevant (a.k.a.~above the line) analyses in {\it Review of Particle Properties} \cite{PDG} show that the partial decay width of $\Delta(1232)$, in the Breit-Wigner sense, is larger than its total decay width.

\begin{table}[h]
    \centering
    \begin{tabular}{|c|c|c|c|}
    \hline
           &  $\Gamma_\mathrm{par}$  & $\Gamma_\mathrm{tot}$  & $\Gamma_\mathrm{par}\,/\,\Gamma_\mathrm{tot}$  \\
     Analysis &   (MeV) &  (MeV) &  (\%) \\
     \hline
       Roenchen \cite{Ronchen2022}  & $100\pm 2$ & $93\pm 1$ & $107.5\pm 2.4$ \\
       \v Svarc \cite{Svarc2014} & $100.0\pm 2.8$ & $98.0\pm 2.2$ & $102.0\pm 3.7$ \\
       Anisovich \cite{Anisovich2012} & $103.2\pm 1.2$ & $99\pm 2$ & $104.2\pm 2.4$ \\
       Cutkosky \cite{Cutkosky1980} & $106\pm 4$ & $100\pm 2$ & $106.0\pm 4.5$ \\
       \hline 
       PDG estimate \cite{PDG} & $100\pm 4$  & $100 \pm ^4 _2$  & $100.0\pm4.5$ \\
       \hline
    \end{tabular}
    \caption{
    The problem: the partial decay width (given by two times the elastic residue magnitude, i.e., $2\,|r|$) of $\Delta(1232)$ is consistently and significantly larger than the total decay width. 
    The statistical average of the ratio is $104.8\pm 2.4$\% 
    }
    \label{tab:Delta1232}
\end{table}

Intriguingly enough, if one takes the PDG estimates for $2 \,|r|$ and $\Gamma$, they are basically the same. As if this problem is totally nonexistent. Therefore, instead of the PDG estimate, we use the statistical average of \mbox{$104.8\pm 2.4$ \%} in the following. 

\begin{figure}[h]
    \centering
    \includegraphics[width=0.47\textwidth]{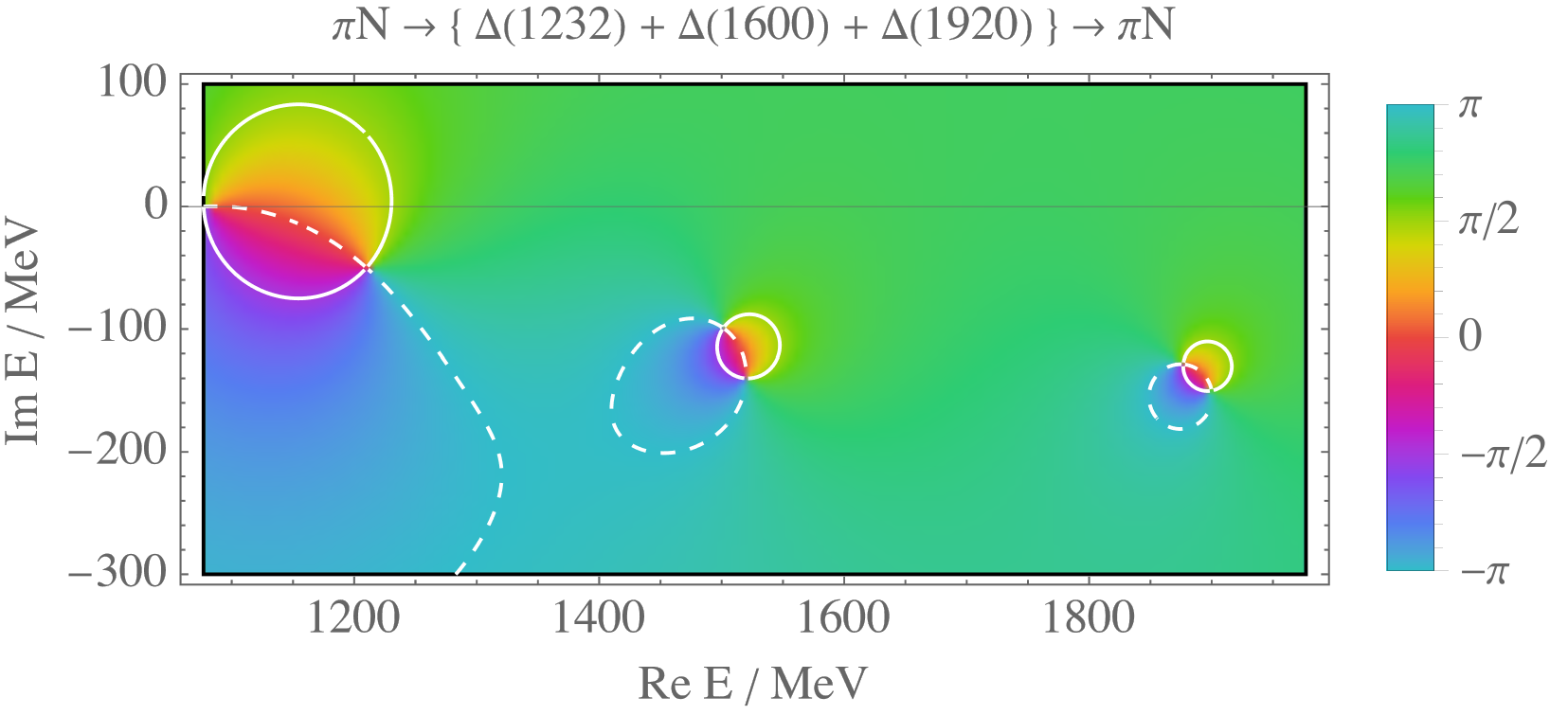}
    \caption{T-matrix phase in the complex energy plane for the $\pi N$ elastic scattering with three included resonances in $\frac{3}{2}^-$ state. The properties of the first resonance remains roughly the same after unitary addition, but for the remaining two, they are drastically changed. 
    }
    \label{fig:delta1232}
\end{figure}

We use Eq.~(\ref{eq:UnitaryAddition}) to produce amplitude $T$ by unitarily multiplying the scattering matrix contributions of the three resonances. The result is given in Fig.~\ref{fig:delta1232}. This is what we find: (1) Two times the magnitude of the elastic residue of $\Delta(1232)$ (the partial decay width in the Breit-Wigner sense) becomes 102~MeV, which is moving in the right direction from initial 100~MeV. (2) For the elastic residue magnitude of $\Delta(1600)$, $\Gamma_\mathrm{par}$ moves from 45 in a single resonance, to 50 in a multiresonance case, in accordance with $50\pm15$ MeV from PDG. (3) The elastic residue of $\Delta(1600)$ we get 12~MeV goes from $-37^\circ$ in single resonance, to a $-92^\circ$ in the multiresonance case, which are again steps in the right direction, towards the PDG estimate of $(-150\pm^{40} _{30})^\circ$. 

 It is important to emphasize this: if our basic model had slightly overestimated the $\Delta(1232)$ partial width compared to its total width, that would not be an issue. We would have just attributed it to the model limitations. However, a consistent discrepancy appears in all relevant analyses. This clearly indicates that there is a real underlying issue.

The real issue here is the assumption that we are obtaining a fundamental resonance property when we extract the pole residue. Even if the pole positions are indeed fundamental resonant properties, their residues depend on all other resonances that have the same quantum numbers, due to scattering matrix unitarity. There is nothing inherently wrong with extraction of complex pole residues, the problem occurs only if one tries to relate them to resonant properties, such is the partial decay width. 

This could be troublesome in meson physics, where the magnitude of the residue often determines the coupling of the resonance to the particular decay channel \cite{PDG}. That coupling will depend on other resonances in the same partial wave. From our excited baryon physics point of view, $\Delta(1232)$ cannot have more than 100 percent probability of decaying to any channel, and therefore we should not interpret the amplitude parameters as the resonance ones. 

In conclusion, this Letter demonstrates that the complex pole residue is not a fundamental property of hadronic resonances. We arrived at this crucial conclusion through a series of compelling findings based on a minimally improved Breit-Wigner formula.

Remarkably, our simple model accurately predicts the residue phases for approximately two-thirds of the most prominent first resonances across $\pi \pi$, $\pi N$, and $KN$ scattering, encompassing $N^*$, $\Delta$, $\Lambda$, and $\Sigma$ baryons, as well as light mesons. This predictive success is particularly striking given the model's fundamental assumption of a simple amplitude zero at the reaction threshold, a simplification that was empirically observed.

Furthermore, our framework provides a resolving explanation for the long-standing conundrum of the $\Delta(1232)$ resonance. Advanced analyses consistently show its partial decay width (as derived from the residue magnitude) exceeding its total width, a seemingly unphysical result. By applying our unitary addition approximation, we successfully reproduce this anomaly, confirming that the residue is not an intrinsic resonance property.

These consistent results across diverse hadronic systems solidify our central assertion: the complex pole residue is a characteristic of the scattering amplitude, profoundly influenced by multi-resonance interference, and cannot be considered a fundamental, isolated property of a resonance. This redefinition necessitates a re-evaluation of how resonance couplings and decay channels are interpreted ensuring a more physically consistent understanding of these excited states.


\end{document}